\documentclass[conference]{IEEEtran}
\IEEEoverridecommandlockouts
\usepackage{cite}
\usepackage{amsmath,amssymb,amsfonts}
\usepackage{algorithmic}
\usepackage{graphicx}
\usepackage{textcomp}
\usepackage{xcolor}
\usepackage{footmisc}
\usepackage{multirow}
\usepackage{url}
\usepackage{tablefootnote}

\def\BibTeX{{\rm B\kern-.05em{\sc i\kern-.025em b}\kern-.08em
    T\kern-.1667em\lower.7ex\hbox{E}\kern-.125emX}}

\begin{document}
\title{Personality and Behavior in Role-based Online Games
\thanks{This work is partly supported by DARPA (grant \#D16AP00115). It does not necessarily reflect the position/policy of the Government; no official endorsement should be inferred. Approved for public release; unlimited distribution.}
}

\author{
\IEEEauthorblockN{Zhao Wang\textsuperscript{1}, Anna Sapienza\textsuperscript{2}, Aron Culotta\textsuperscript{1}, Emilio Ferrara\textsuperscript{2}}
\IEEEauthorblockA{\textit{\textsuperscript{1}Department of Computer Science, Illinois Institute of Technology, Chicago, USA} \\
\textit{\textsuperscript{2}Information Sciences Institute, University of Southern California, Los Angeles, USA} \\
\{zwang185, aculotta\}@hawk.iit.edu, \{annas, ferrarae\}@isi.edu}}





\IEEEpubid{\begin{minipage}{\textwidth}\ \\[12pt]
978-1-7281-1884-0/19/\$31.00 \copyright 2019 IEEE
\end{minipage}}

\maketitle

\begin{abstract}
Both offline and online human behaviors are affected by personality. Of special interests are online games, where players have to impersonate specific roles and their behaviors are extensively tracked by the game. In this paper, we propose to study the relationship between players' personality and game behavior in League of Legends (LoL), one of the most popular Multiplayer Online Battle Arena (MOBA) games. We use linear mixed effects (LME) models to describe relationships between players' personality traits (measured by the Five Factor Model) and two major aspects of the game: the impersonated roles and in-game actions. On the one hand, we study relationships within the game environment by modeling role attributes from match behaviors and vice versa. On the other hand, we analyze the relationship between a player's five personality traits and their game behavior by showing significant correlations between each personality trait and the set of corresponding behaviors. Our findings suggest that personality and behavior are highly entangled and provide a new perspective to understand how personality can affect behavior in role-based online games.
\end{abstract}


\section{Introduction}
\label{sec:Introduction}
Numerous studies from cognitive science and social science have shown strong connections between personality and human behavior~\cite{PsyOps, Narnia2014role, McCreery2012}. As an important part of human activity, digital footprints are found to be closely related to user personality. For instance, extroverts and emotionally stable people are shown to be popular and influential on Twitter~\cite{Quercia@2011}. Meanwhile, personality can be successfully inferred from user-generated contents in social media (e.g., Facebook, Twitter, and YouTube)~\cite{Farnadi@2016}. As a modern entertainment, online games are becoming increasingly popular, with an average time spent in playing of more than 20 hours per week~\cite{Narnia2014role}. Given the popularity achieved, these games constitute a desirable source of data, providing valuable opportunities for researchers to investigate potential connections between personality and behavior. Of special interests are Multi-players Online Battle Arena (MOBA) games, which include several social aspects that could be influenced by personality, such as impersonating a role and cooperating with teammates.


Here, we study one of the most popular MOBA games: League of Legends (LoL). In LoL game scenario, each player selects a specific role by impersonating a champion, i.e. a character of the game, to conduct a series of actions during the match. Each role is designed from the prototype of a champion in the fairy tale and characterized by a set of unique abilities~\cite{LolArt}. As humans have personalities that describe their stable behavioral patterns, champions also have attributes that characterize their abilities in the game scenario. Players with different personalities would have different preferences for specific roles, which subsequently lead to different game behaviors.



Numerous researches have been conducted to investigate the relationship between personality and game behavior by profiling players' personality through their actions in the game~\cite{Lankveld@6032007}, by conducting surveys to track both personality and self-reported in-game behaviors~\cite{Narnia2014role}, and by taking demographic effects (e.g., gender and age) into account~\cite{Yee:2006:DMD:1159982.1159988, GOLDBERG1998393, Griffiths2004}. However, these works do not consider the different roles chosen by players, and often rely on self-reported game actions, which could lead to biases and inaccuracies in the data.  

By extending the previous research in this field, we aim at both studying how players' personalities are related to their preferences in selecting game characters and understanding how personalities influence game behavior. We first conduct surveys to collect the BIG-5 personality traits of players, as discussed in Section~\ref{sec:data}. Then, we combine data of players' impersonated characters and actions provided by the official Riot Games API.\footnote{\label{lol_api}\url{https://developer.riotgames.com/}} Our main goal is to understand: first, how a player's impersonated role and their match behavior are related; and second, whether knowing a player's role and match behavior helps to understand their personality. To this aim, we apply Linear Mixed Effects (LME) models (Section~\ref{sec:method}) to find the underlying relations between LoL champions (i.e. game characters) and the actions performed during the match (Section~\ref{sec:champ_match}). In Section~\ref{sec:personality}, we analyze players' personality traits according to the attributes of impersonated champions and match behaviors. Finally, in Section~\ref{sec:futureWork}, we report the conclusions, limitations, and future work.

\section{Related Work}
\label{sec:relatedWork}
Previous studies focused on exploring the relationship between personality and game behaviors. In~\cite{Narnia2014role}, the authors conduct a study based on 205 players of {\it World of Warcraft} and find that personality traits strongly influence game behaviors. However, this work collects game behaviors through surveys, which might suffer from inadequacy and reliability. Authors in~\cite{Lankveld@6032007} develop a framework to automatically collect playing behaviors in {\it Neverwinter Nights}, and then use regression analysis to profile 44 players through the collected behaviors. Results suggest significant correlations between personality and behavior in role-playing game, nevertheless, this work does not take into account possible effects caused by different roles. Additionally, players with different personalities are shown to differ in: motivations for playing~\cite{Suler2004}, preference of game genres~\cite{borders2012relationship}, selection of game characters~\cite{Delhove2018}, and engagement of match behaviors~\cite{Yee2011IntrovertedE}. Demographic variables (e.g., gender, age) were also found to influence these aspects of the game~\cite{Yee:2006:DMD:1159982.1159988}. 


A great amount of work has been also devoted to study the connections between personality and offline behavior. However, it is not clear how to transfer the related findings to have better insights into online behavior. On the one hand, some studies found consistencies between personality and offline behaviors as well as personality and online behaviors~\cite{Graham2013, Johnson:2010:PMV:1952222.1952281,Lankveld2009,Peng2008}. For instance, {\it Agreeableness} and {\it Extraversion} are shown to be connected to playing motivations. On the other hand, there are also works reporting inconsistencies between personality traits and behaviors, such as inconsistency of {\it Conscientiousness} and motivations in~\cite{Park2011}. 

In the present work, we aim at shedding light on the underlying connections between personality and game behavior by taking big-5 personality traits, role attributes, actions during matches, and demographic factors all into account.


\section{Data}
\label{sec:data}
\begin{figure*}[t!]
	\centering
	\includegraphics[width=\textwidth]{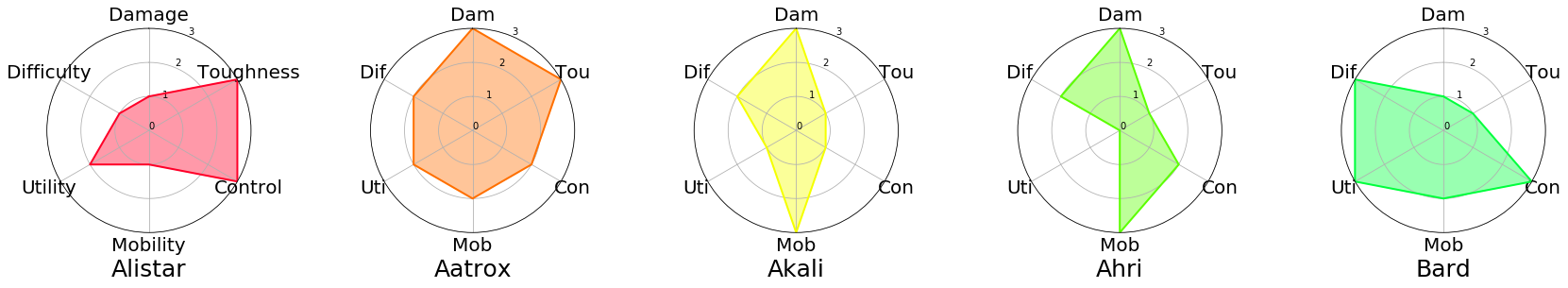}
	\caption{ Example of radar plots for five different champions (i.e., Alistar, Aatrox, Akali, Ahri, Bard), each defined by six attributes (we use the first three letters as abbreviations of attribute names). We observe that every champion has its weaknesses and strengths regarding the six attributes and no champion is designed to be perfect~\cite{Anna2017ICDMW}, so that champions in a team need to cooperate to win.}
	\label{fig.champ}
\end{figure*}

\begin{figure*}[t!]
	\centering
	\includegraphics[width=\textwidth]{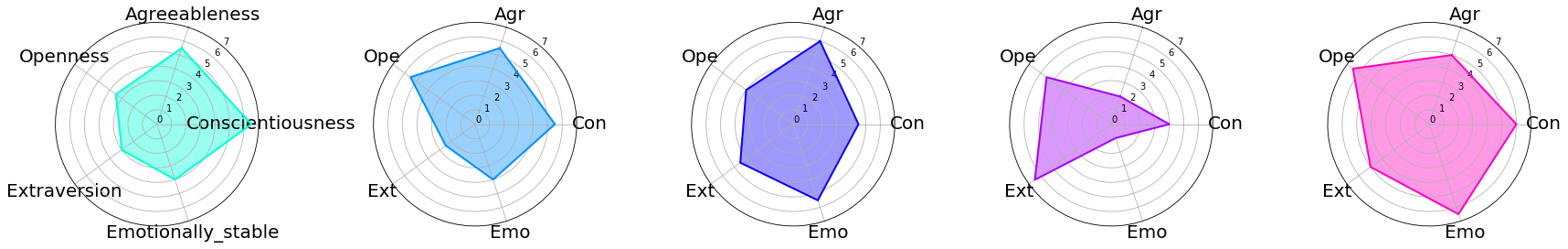}
	\caption{Example of radar plots for big-5 personality traits of five different players (we use the first three letters as abbreviations of personality trait names). Players' big-5 personality traits are analogous to champion attributes: each player has its weaknesses and strengths regarding the five dimensions of personality traits. Players with different personalities will have different preferences for champions.} 
	\label{fig.big5}
\end{figure*}

In this paper, we study the relationship between personality and game behavior of {\it League of Legends} (LoL) players. We first collect personality traits data through a survey containing personality test and demographic questions.\footnote{An anonymized version of the dataset used in this study is available upon request to the contact author.} Then, for each player participating in the survey, we collect their game information (e.g., selected characters and match history) via the official Riot Games API.\footref{lol_api}

\subsection{League of Legends (LoL)}
As one of the most popular MOBA games, LoL was first released in 2009 and updated for nine seasons by the end of 2018~\cite{LOLESPORTS2018}. It attracts millions of players of all ages, nationalities, and occupations. What's more, LoL is reported to have over 80 million active players per month and over 27 million players every single day~\cite{Sheer2014}.

LoL offers a variety of options allowing players to interact with each other in a virtual environment full of activities and diverse game modes~\cite{LoLguide}. There are 141 unique characters (i.e. champions in LoL), each defined by a set of special abilities. We summarize these abilities into six main attributes to characterize each champion:\footnote{ Champions and attributes definitions (copy the link into your browser):\\ \url{https://leagueoflegends.fandom.com/wiki/List_of_champions/Ratings}}
\begin{itemize}
    \item Damage: {\it ``ability to deal damage''}.
    \item Toughness: {\it ``ability to survive being focused''}.
    \item Mobility: {\it ``ability to move quickly, blink or dash''}.
    \item Utility: {\it ``ability to grant beneficial effects on allies, or provide vision''}.
    \item Control: {\it ``ability to disable enemies''}.
    \item Difficulty: {\it ``a champion's mechanical difficulty''}.
\end{itemize}
Each attribute has a rating that ranges from 0 to 3, with 0 representing the weakest level and 3 representing the strongest. Fig.\ref{fig.champ} shows examples of five different champions' attributes. A single match in the Summoner's Rift (i.e. the most popular game mode of LoL) requires two teams of five players that compete to destroy the enemy base, and each player controls one of the 141 champions. Therefore, the champion selection and playing style are critical to the outcome of a match, both of which can be strongly influenced by players' personalities.


\subsection{Survey}
We conducted a survey to collect players' personality traits and demographic information. A sample of our survey is available online.\footnote{\url{https://www.psytoolkit.org/cgi-bin/psy2.5.1/survey?s=LUPWZ}} The questionnaire contains three categories of questions: personality test, demographic questions, and player identification questions. 

Among various personality models, the BIG-5 framework has emerged as the most widely accepted model to study personality traits~\cite{Gosling2003big-five, Digman1990, Cattell1996}. In our setting, we choose {\it BIG-5 Ten Item Personality Inventory} (TIPI) as our personality questionnaire, which is brief but proven to be reliable~\cite{Gosling2003big-five}. We obtain the TIPI questionnaire from PsyToolkit:\footnote{\url{https://www.psytoolkit.org/survey-library/big5-tipi.html}} a free survey library that provides cognitive-psychological tests~\cite{Stoet2010, Stoet2017}. Each of the five personality dimensions, i.e. {\it agreeableness, conscientiousness, emotional stability,\footnote{Also known as an antonym of neuroticism~\cite{Gosling2003big-five}.}, extraversion, and openness} is measured according to two questions of the TIPI test~\cite{Norman@1963}. Participants are asked to rate their answers to each question on a seven-point scale ranging from 1 (i.e. strongly disagree) to 7 (i.e. strongly agree). TABLE~\ref{tab.std} shows the mean and standard deviation for personality scores of valid participants. Fig.\ref{fig.big5} shows examples of five different players' personality traits. Detailed definitions of the big-5 personality traits are provided in Section~\ref{model_big5}.

\begin{table}[t!]
    \caption{ Mean and standard deviation for big-5 personalty traits of 811 valid participants}
    \centering
    \begin{tabular}{| c | c | c | c | c | c |} 
     \hline
      {} & \textbf{Agr} & \textbf{Con} & \textbf{Emo} & \textbf{Ext} & \textbf{Ope}\\
     \hline
     \textbf{Mean} & 4.42 & 4.70 & 4.60 & 3.53 & 4.95\\
     \hline
     \textbf{Std} & 1.24 & 1.31 & 1.46 & 1.52 & 1.20\\
     \hline
     \multicolumn{6}{l}{$^{\mathrm{a}}${\it We use first 3 letters to represent trait names.}}

    \end{tabular}
    
    \label{tab.std}
    \vspace{-0.5cm}
\end{table}

To control for the potential effects that demographic factors have on personality~\cite{Narnia2014role, GOLDBERG1998393, Griffiths2004, Yee:2006:DMD:1159982.1159988}, we also ask for players' demographic information (i.e. age, gender, region of the game) and translate the survey into several languages (i.e. Korean, Japanese, Turkish, Spanish, and Chinese) to reach out players of different nationalities. 

Moreover, we ask the following identification questions: the summoner name\footnote{\url{https://na.leagueoflegends.com/en/game-info/summoners/}} (i.e. the unique identification name of each player); three favorite champions\footnote{\url{https://na.leagueoflegends.com/en/game-info/champions/}}  (i.e. characters a player prefers to impersonate during the game); and the highest champion level. These questions are used to verify that the participant is an actual LoL player. Finally, we also record players' response time for each question. Specifically, for the 10 personality questions, the mean response time is 50s with a standard deviation of 25s.

We post the survey on several platforms, such as LoL forums,\footnote{\url{https://boards.na.leagueoflegends.com/en/}} popular websites (e.g., Reddit), and Amazon Mechanical Turk (AMT). However, despite the great interest shown by volunteers when the survey was released, the overall attention dramatically decayed after a few days. We then focused on AMT, where we boost participation by providing a 0.5\$ reward for each valid response. We conducted the survey on AMT for three months (June to September 2018).

We use the following criteria to validate the authenticity of participants. First, we remove duplicate responses having the same summoner name or AMT worker ID. Second, we check if a participant's answers to the identification questions match the information we get from the official API of LoL. Since the identification information is only visible to a player on their own account, we can identify false LoL players through this process. Finally, we require a participant to have played at least 10 matches to filter possible biases introduced by players who only play the game a few times and have insufficient information to assess their game behavior.

At the end of this procedure, we finally get 811 valid responses out of 2785 in total. TABLE~\ref{tab.demographic} shows a summary of the demographic statistics of verified survey participants. Note that, since we are focusing on AMTs, the participants' demographics are affected by AMT demographics.

\begin{table}[t!]
    \caption{Demographic statistics of 811 valid survey participants}
    \centering
    \begin{tabular}{| l | l | l |} 
     \hline
      \textbf{Gender} & \textbf{Age} & \textbf{Region}\\
     \hline
      M:682 & 13$\sim$20: 195 & NA1 (North America): 580\\
      F:129 & 21$\sim$30: 496 & EUW1 (EU West): 107\\
       & 31$\sim$40: 100 & others: 124\\
       & $>$40: 20 & \\
     \hline
    
    \multicolumn{3}{l}{$^{\mathrm{a}}${\it Minimum age of a LoL player is required to be 13\tablefootnote{\url{https://na.leagueoflegends.com/en/legal/termsofuse}}}} \\
    
    \multicolumn{3}{l}{$^{\mathrm{b}}${\it Regions are defined by server locations (check full list of regions here\tablefootnote{\url{https://leagueoflegends.fandom.com/wiki/Servers}})}} \\
    
    \end{tabular}
    \label{tab.demographic}
    \vspace{-0.5cm}
\end{table}

\subsection{Game Behavior Data}\label{subsec:champ}
The Riot Games API provides easy access to LoL game data in a secure and reliable way. For the 811 valid players, we utilize the API to collect the record of champions a player has impersonated, the timeline of their matches, and the specific actions performed during each match (e.g., the number of kills, deaths, and assists). The statistics of collected champions and matches are shown in the histograms of Fig.~\ref{fig.pearson_match_champ}.




\section{Methods}
\label{sec:method}
In this section, we first explain the representative features of players. Then, we describe the technical background of linear mixed effects models and show how we apply them to explore potential relationships between personality and game behavior.

\subsection{Features}
We extracted four sets of features from the collected data: demographic factors, big-5 personality traits, champion features, and match behaviors.

\textbf{\textit{Demographic factors:}} age, gender, and region.
    
\textbf{\textit{Big-5 personality traits:}} agreeableness, conscientiousness, emotional stability, extraversion, and openness.
    
\textbf{\textit{Champion features:}} 
we consider the following features to describe the champions impersonated by each player:

\begin{itemize}
    \item n\_champs: the total number of unique champions used by each player, as the horizontal axis shows in Fig.\ref{fig.pearson_match_champ};
    \item n\_level6: the number of champions ranked no lower than level six (champion level is determined by the frequency and proficiency of its usage by a player); This is a complementary feature of n\_champs, which helps to identify a player's performance level and characterize the expertise of a player's skill.
    \item Average attribute ratings of a player's three favorite champions (used to measure a player's champion preference);
    \item Weighted attribute ratings of a player's top three ranked champions (weight by the number of matches being played with each specific champion), which is used to measure a player's champion skill. 
\end{itemize}
    
\textbf{\textit{Match behavior:}} We select the actions of top three ranked champions of a player to represent their overall match behavior. 
The official API records more than 100 behavioral features for each match, including kills, deaths, and assists (the complete set of features can be found at official API\footref{lol_api} webpage). However, some features are either not representative of a player's game behavior (e.g., visionScore) or correlated with other features (e.g., kills, doubleKills, tripleKills, and quadraKills are highly correlated), or not recorded for most players (e.g., combatPlayerScore). To focus on features that are most relevant to the playing behavior, we remove non-representative features and highly correlated features, and only keep features that are recorded for most players. By doing so, we extract 11 significant features to describe important aspects of the playing behavior: the total number of matches (n\_matches, as the vertical axis shows in Fig.\ref{fig.pearson_match_champ}), number of kills, number of deaths, number of assists, winning rate, average match duration, gold earned, gold spent, total damage dealt, total damage taken, total heal. 


\subsection{Linear Mixed Effects Models (LME)}
Previous studies have shown that both playing behaviors and personality traits are affected by demographics (e.g., gender, age)~\cite{Narnia2014role, GOLDBERG1998393}. For example, female players are more engaged in assisting behaviors than male players, and young players tend to be less emotionally stable than old ones~\cite{Yee2011IntrovertedE}. However, mixing demographic effects with playing behaviors and personality traits would impede the identification of possible relationships between personality and game behavior. 


To deal with this issue, we apply LME models, which allow us to consider both fixed and random effects. Fixed effects refer to variations that could be explained by the independent variables (like in linear models) and random effects refer to variations that could not be explained by independent variables. By following the syntax provided in~\cite{JSSv067i01}, we can write the equations for mixed effects models as follows:
\begin{equation}
    \text{outcome} \sim 1 + \text{fixed effects} + \left(\text{random effects} | \text{group}\right)\;
\end{equation}
where an outcome (dependent) variable is explained by using an intercept equal to 1, one or more fixed effects, and one or more random effects allowing for variations between groups (e.g., gender group: male, female). We use the R implementation of {\it lme4} to perform our linear mixed effects analysis in all the experiments~\cite{Bates2011, RCoreTeam}.

To model each of the three aspects: big-5 personality traits, champion attributes, and match behavioral features, we refer to the one being modeled as the outcome variable, while the remaining aspects are considered to be the candidate fixed effects (which are constant across demographic groups~\cite{gelman2005}). Finally, we control for different demographic groups by considering them as random effects. 

To assess the relative fits of the models, on the one hand, we check their linearity, homoscedasticity, and normality of the residuals. On the other hand, we compute the Bayesian Information Criterion (BIC)~\cite{schwarz1978}, which accounts for both the complexity and likelihood of a model. For each outcome variable, we start with a null hypothesis which only explains the outcome by the intercept and random effects of gender, age, and region. Then, we incrementally add fixed effects and select those that decrease the BIC score of the model. Finally, we use ANOVA tests to compare the differences between the null model and the one with the lowest BIC score. According to the interpretation table of BIC score differences in~\cite{Robert1995BIC}, we then determine the significance of each model and report the one that is the most significant (i.e. the model whose BIC score difference with the null hypothesis model is greater than 10 and has the lowest BIC score).

\section{Relationship between Role and Match Behavior}
\label{sec:champ_match}
\begin{table*}[t!]
    \caption{ Inferred champion attributes from match behavior}
    \centering
    \begin{tabular}{| c | c | c | c | c | c | c | c |}
     \hline
      \multicolumn{2}{|c|}{} & \textbf{Control} & \textbf{Damage} & \textbf{Difficulty} & \textbf{Mobility}  & \textbf{Toughness} & \textbf{Utility}\\
     \hline
     \multirow{3}{*}{\textbf{Random Effects}} & gender &  &  & \checkmark & \checkmark &  &  \\
     & age &  & \checkmark & \checkmark & \checkmark &  & \checkmark \\
     & region  &  & \checkmark & \checkmark & \checkmark &  &  \\
     \hline
     \multirow{11}{*}{\textbf{Fixed Effects}} & kills & -0.34 & 0.79 & 1.05 & 0.81 &  & -0.55 \\
     & assists & 0.62 & -0.91 & 0.11 & -0.45 & 0.24 & 0.53 \\
     & deaths & -0.25 & 0.75 &  &  & -1.56 & 0.45 \\
     & win & 0.31 & -0.49 &  & -0.17 &  & 0.41 \\
     & average duration & 0.38 & -0.41 & 0.51 &  &  & 0.30 \\
     & champLevel & 0.61 &  & -0.18 &  & 0.71 &  \\
     & goldEarned & -1.18 & 1.35 & -1.56 &  & -1.27 & -1.04 \\
     & goldSpent &  &  & 0.83 & 0.78 & -0.54 &  \\
     & totalDamageDealt &  &  &  & -0.29 &  &  \\
     & totalDamageTaken & 0.19 & -0.75 &  &  & 2.00 & -0.49 \\
     & totalHeal & -0.20 & -0.26 &  & 0.21 & -1.04 & 0.44 \\
     \hline
     \multicolumn{8}{l}{$^{\mathrm{a}}$ {\it Notes apply to all the following tables: }}\\
     \multicolumn{8}{l}{\it (1) Results should be interpreted by column; Feature values are pre-processed by applying min-max normalization;} \\
     \multicolumn{8}{l}{\it (2) For random effects, we use check marks to indicate which plays an important role;} \\
     \multicolumn{8}{l}{\it (3) For fixed effects, we only show those with significance p$<$0.05;}\\
     
    \end{tabular}
    \label{tab.champ_LME}
\end{table*}

\begin{figure}[t!]
	\centering
	\includegraphics[width=0.9\columnwidth]{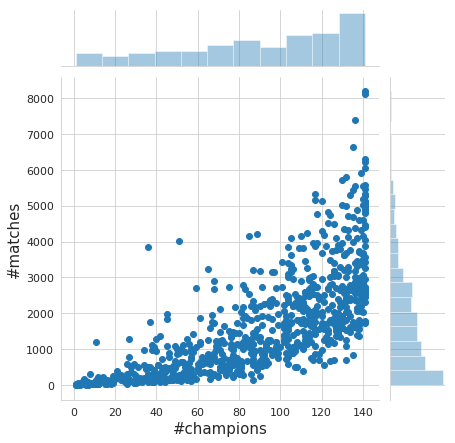}
	\caption{Relation between number of champions and matches for 811 players, each point represents one player.}
	\label{fig.pearson_match_champ}
\end{figure}

Players choose champions to conduct a series of actions during the game and the champion's abilities would directly affect players' match behaviors~\cite{Bean2014}. We start by showing the relationship between champion usage and matches. Then, we model champion attributes from match behavior as well as infer specific match behavior according to champion attributes. 

\subsection{Relationship between Champions and Matches}


To understand how the number of different champions used by players varies with the number of matches being played, we first compute the Spearman correlations between these two components and results show that they are strongly correlated (0.813 with p\_value $<$ 0.01). Fig.~\ref{fig.pearson_match_champ} shows the scatter plot of the number of champions and matches for 811 valid players. Histograms on two axes represent the density of champions and matches in each interval. 

We can observe that the number of different champions each player tries in their gaming history is positively correlated with the number of matches s/he plays. Moreover, this relation follows an exponential trend, which suggests that at the start of their gaming history, players take few champions to learn how to play the game, and later on, they are more inclined to explore new characters. 


Despite the overall trend, we find cases of players who play most of their matches with a small number of champions, and cases of players who explore the use of different champions in the early stages of their gaming history. The former case is indicative of players who tend to have extremely limited interest in champions. The latter case is instead indicative of players with a broad interest in exploring new options.

\subsection{Model Champion Attributes from Match Behavior}\label{model_champ4.2}

Each champion has specific abilities and these abilities would subsequently affect match behaviors, thus we are interested in understanding how do match behaviors reflect champion abilities. We focus on the six main attributes that characterize champion abilities: control, damage, difficulty, mobility, toughness, and utility. We then formulate this problem as LME models for champion attributes. Here, each attribute is modeled with the 11 match features as candidate fixed effects and three demographic factors as candidate random effects. 


TABLE~\ref{tab.champ_LME} shows the relationship between champion attributes and the set of most relevant match behaviors. We observe that: (1) demographic features play an important role as random effects to model attributes of damage, difficulty, mobility, and utility; (2) the 11 match behavior features are reflective of champion attributes, which can be backed up by the statistics of matched games from Champion.gg,\footnote{\label{champgg}\url{https://champion.gg}} a website that provides in-depth and accurate statistics about the overall performance of each champion in LoL. Take the attribute ``Damage'' for an example, TABLE~\ref{tab.champ_LME} shows that a large number of kills and deaths but a small number of assists are indicative of high damage rating of the champion. According to the statistics in Champion.gg, we find that the top five champions conducting a large number of kills and a small number of assists are indeed champions with high damage ratings. For instance, the champion {\it Quinn}, having damage rating of 3 (the highest level), being ranked 2nd for average number of kills, and being in the lowest ranks for number of assists.\footnote{\url{https://champion.gg/champion/Quinn/Top?league=gold}}

\subsection{Infer Match Behavior according to Champion Attributes}

\begin{table*}[t!]
    \caption{ Inferred match behaviors according to champion attributes}
    
    \centering
    \begin{tabular}{| c | c | c | c | c | c |} 
     \hline
      \multicolumn{2}{|c|}{} & \textbf{Kills} & \textbf{Assists} & \textbf{TotalDamageTaken} & \textbf{TotalHeal}\\
     \hline
     \multirow{3}{*}{\textbf{Random Effects}}
     & gender & \checkmark &  & \checkmark &  \\
     & age & \checkmark & \checkmark & \checkmark & \checkmark \\
     & region  & \checkmark & \checkmark & \checkmark & \checkmark \\
     \hline
     \multirow{6}{*}{\textbf{Fixed Effects}}
     & control & -0.09 &  & -0.17 & -0.16 \\
     & damage & 0.11 & -0.16 &  & -0.15 \\
     & difficulty &  &  &  & -0.05 \\
     & mobility &  & -0.08 &  &  \\
     & toughness &  & -0.10 & 0.25 & 0.09 \\
     & utility & -0.06 & 0.08 & -0.07 & 0.05 \\
     \hline
    \end{tabular}
    
    \label{tab.match_LME}
\end{table*}

After modeling champion attributes from match behavior, we are then interested in understanding how the choice of certain champions affect playing behaviors in the game. We answer this question by fitting a LME model for each match feature. In this model, we use the six champion attributes as candidate fixed effects and the three demographic factors as candidate random effects. 


We fit one LME model for each of the 11 match features and identified significant LME models for kills, assists, totalDamageTaken and totalHeal. TABLE~\ref{tab.match_LME} shows the relationship between match features and champion attributes indicated by corresponding LME models. Results show that demographic features also play an important role in understanding how champion choice affects match behaviors. Moreover, the partial combination of champion attributes is indicative of specific match behaviors. For instance, champions with low control rating, high damage rating, and low utility rating are expected to have more kills; champions with low damage, low mobility, low toughness, and high utility ratings are expected to have more assists. These relationships also reflect the statistics of matched games from Champion.gg.\footref{champgg} In particular, the champion {\it Master Yi} has the following attribute ratings: {\it Damage:3, Control:0, Utility:0, Mobility:2, Toughness:1, Difficulty:1}, and is ranked 5th on the basis of their number of kills. 

Given these results, we found that there exists an intrinsic relation between impersonated champions and their actions during the game. We are now interested in making a step further to check whether a relation exists between the impersonated champion and personality. This will allow us to understand how personality influences a player's choice of roles as well as to directly link their personality with the actions performed during the game.

\section{Understanding Personality from Game Behavior}
\label{sec:personality}
In this section, we first show an interesting relationship between players' favorite champions and most skilled champions. Then, we examine how do big-5 personality traits align with champion choices and corresponding match behaviors in the online game environment.

\begin{table*}[h!]
    \caption{ Understanding big-5 personality traits from champion attributes and match behavior}
    \centering
    \begin{tabular}{| c | c | c | c | c | c | c |} 
     \hline
      \multicolumn{2}{|c|}{} & \textbf{Agreeableness} & \textbf{Conscientiousness} & \textbf{Emotional stability} & \textbf{Extraversion}  & \textbf{Openness}\\
     \hline
     \multirow{3}{*}{\textbf{Random Effects}}
     & gender & \checkmark &  & \checkmark & \checkmark & \checkmark\\
     & age &  & \checkmark & \checkmark & \checkmark & \checkmark\\
     & region & \checkmark &  &  & \checkmark & \\
     \hline
     
     \multirow{14}{*}{\textbf{Fixed Effects}} & control & -1.34 & 1.03 & 3.28 & -0.72 & 1.59 \\
     & damage & -1.04 & -3.82 &  & 1.47 & 1.45 \\
     & difficulty & -1.47 &  & -1.49 & -1.49 & -2.34 \\
     & mobility & 0.76 &  & -2.06 & 4.35 & -4.30 \\
     & toughness & -1.28 & -2.46 & -1.61 & -5.24 & -6.37 \\
     & utility & -0.54 & 1.49 & 1.27 & -4.88 & 2.35 \\
     &  &  &  &  &  &  \\
     
     & kills & 9.22 & -5.26 & 8.71 & 17.28 & 5.64 \\
     & deaths & -8.10 &  & -4.04 & -11.9 & -4.33 \\
     & assists & -2.92 &  & -5.63 & -4.36 & -7.43 \\
     & win & 15.6 & 2.56 & 3.05 & 11.92 & 15.31 \\
     & average duration & -0.01 & -0.01 & 0.09 & -0.03 & 2.53 \\
     & n\_level6 & 0.06 & -0.04 & 0.07 & 0.12 & -1.73 \\
     \hline
     
    \end{tabular}
    
    \label{tab.big5LME}
\end{table*}

\subsection{Favorite vs Skilled Champions}
To understand whether players are good at playing with their favorite champions, we compare players' favorite three champions (collected in the survey) with their top three ranked champions (collected via the official API). Results show that for 83\% of players, at least one champion is both their favorite and skilled, while for 17\% players, their favorite champions are totally different from the skilled ones. This finding suggests that players' skilled champions are not always exactly their favorite ones. In particular, players might have the tendency of using champions that are easy to control and win despite personal preference.


\subsection{Understanding Personality from Game Behavior} \label{model_big5}
As personality could affect champion choices and champion abilities would subsequently influence playing behaviors. Here, we focus on understanding personality traits from impersonated roles and corresponding playing behaviors, which would corroborate the previous results showing the existence of associations between players' personality and playing behaviors in online game  environments~\cite{Narnia2014role, Lankveld2009}. 

To explore such associations, we formulate the problem as LME modeling tasks for the big-5 personality traits. Specifically, with champion attributes and match features as candidate fixed effects and demographic factors as candidate random effects, our goal is to explore relationships between personality traits and game aspects (including champion choices and match behaviors). 

We first conduct the analysis by using all of 811 samples, but results show weak correlations between personality and game aspects, which might be caused by either noisy samples or neutral personality traits (e.g., some players show neutral ratings for agreeableness). On the one hand, to limit possible noisy samples, which is a general problem in survey-based studies, we follow the criteria of response time reported in~\cite{Lankveld2009} and limit our samples to those whose response time for the personality test is greater than two minutes. On the other hand, we label each trait according to its first and third quantiles, and only keep those whose personality score is smaller than the first quantile (label as low) or greater than the third quantile (label as high). This allows us to study players with obvious differences in personalities and avoid confusions introduced by those with neutral personality traits. 



We then re-fit our LME models on the obtained sub-samples regarding each trait and get more consistent and significant relationships. TABLE~\ref{tab.big5LME} illustrates the correlations between each personality trait and selected game aspects (including champion attributes and match features). We will explain the relationship
regarding each personality trait in turn below.

Agreeableness is the tendency of getting along with others in pleasant, satisfying relationships~\cite{Norman@1963}. TABLE~\ref{tab.big5LME} shows that the relationship between agreeableness and game behavior is mainly affected by gender and region. In our samples, males show lower agreeableness than females; players in region EUW1 (i.e., European West) show lower agreeableness than players from other regions. Results also show that agreeableness is negatively correlated with utility in LoL, which is different from the case in a real-life scenario~\cite{Farnadi@2016}. Additionally, since people with high agreeableness are expected to cooperate well with teammates, it is reasonable to find that they are characterized by a strong killing behavior and fewer deaths, which is related to good cooperation with teammates and a key component to achieve a high winning rate.

Conscientiousness is the personality trait of being careful and having the desire to positively fulfill a task~\cite{Poropat2009, Yee2011IntrovertedE}. According to TABLE~\ref{tab.big5LME}, the relationship between conscientiousness and game behavior is mainly affected by age, where older players tend to have higher conscientiousness than younger ones. Results also show that conscientious players prefer champions with good control ability which allow for better management in dangerous situations. The nature of being careful and keep things under control make players more likely to win. We also notice that conscientious players have a relatively small number of skilled champions, which suggests that conscientious players mainly focus on specific champions and are not exploring new possibilities of different champions.

Emotional stability is the ability of calmly handling difficult situations~\cite{Peng2008, Gosling2003big-five}. Results in TABLE~\ref{tab.big5LME} show that the relationship between emotional stability and game behavior is mainly affected by gender and age. According to our samples, males tend to be more emotionally stable than females; players between $21\sim30$ are reported to be more emotionally stable than others. Moreover, emotionally stable players are able to take champions with high control and utility ratings as well as performing more kills while keeping a low death rate. Additionally, we observe that the average match durations for emotionally stable players are relatively longer. All these behaviors correspond to their ability of being stable and productive in tough situations.

Extraversion is the trait of being outgoing and social~\cite{PsyOps, Narnia2014role}. Its relationship with game behavior is affected by gender, age, and region. In our sample, males show lower extraversion than females, older players tend to be more extrovert than younger ones, and players in EUW1 show lower extraversion than those in other regions. TABLE~\ref{tab.big5LME} also indicates that extrovert players prefer champions with high damage and mobility ratings, and thus are prone to kill more and have a low death rate, which leads to a high winning rate. It also shows that extrovert players have a relatively large number of expert champions, which means these players have a broad interest in trying different champions. These behaviors support the outgoing, social and optimistic aspects of being extrovert.

Openness is being characterized by high curiosity and creativity. Its relationship with game behavior is slightly affected by gender and age. Players with high openness tend to choose champions with high control, damage, and utility ratings, leading to relatively more killing behaviors, fewer deaths, and thus higher winning rate. However, we notice that players with high openness do not have a large number of assisting behaviors and expert champions, which differs from the real-life behavior of people with the nature of being open.

\section{Conclusions and Limitations}
\label{sec:futureWork}
In this paper, we explore the relationship between personality and behavior in LoL. We collect data of players' in-game behavior from the official Riot Games API and adopt the Five-Factor Model to get survey-based personality traits of LoL players. We applied linear mixed effects models to fit our data and describe the entangled relationships between personality and playing behaviors while taking demographic random effects into consideration. 

First, we highlight the exponential relationship between the number of champions being used and matches being played. We then study the relationship between champions and match behaviors by modeling champion attributes and playing behaviors from each other. For instance, results show that champions with low control rating and utility rating but high damage rating are expected to have more killing behaviors. Second, we investigate the relations between a player's personality and game aspects (including impersonated champions and corresponding match behaviors). Results show significant associations among these factors. For example, we observed that conscientious players prefer champions with good control ability to face dangerous situations, and that the nature of being careful makes these players more likely to win.

Future work will be devoted to overcoming some limitations of the current study. We plan to collect more data as the current analysis is based on the data of 811 players. Despite this amount of participants is sufficient from the psychological perspective of personality analysis, it limits us from applying complex machine learning models as well as conducting more nuanced analysis (e.g., considering the change of match statistics when champions fill different roles, comparing performance of low-rank and high-rank players). A bigger and more diverse dataset will also allow overcoming problems due to misbehavior of survey participants. LoL is a worldwide game, but we mainly spread our personality questionnaires in US websites and Amazon Mechanical Turk. Thus the demographics of participants are not universally balanced. For future work, we will improve the recruitment of participants to have broader samples.

In conclusion, we firmly believe that the results of our work could benefit both companies and players. The former could develop customized game characters or carry out personalized recommendations based on players' historical match behaviors. The latter would have access to a way of understanding how their personalities affect game behaviors, thus having insights of how to successfully assemble teams and develop reasonable strategies during the match. 


\bibliographystyle{unsrt}
\bibliography{reference}

\begin{thebibliography}{10}

\bibitem{PsyOps}
Shoshannah Tekofsky, Pieter Spronck, Aske Plaat, and Plaat@uvt.
\newblock Psyops: Personality assessment through gaming behavior.
\newblock In {\em the 8th International Conference on the Foundations of
  Digital Games}, 2013.

\bibitem{Narnia2014role}
Narnia~C. Worth and Angela~S. Book.
\newblock Personality and behavior in a massively multiplayer online
  role-playing game.
\newblock {\em Computers in Human Behavior}, 38:322 -- 330, 2014.

\bibitem{McCreery2012}
P.G. Schrader Randy~Boone Michael P.~McCreery, S. Kathleen~Krach.
\newblock Defining the virtual self: Personality, behavior, and the psychology
  of embodiment.
\newblock {\em Computers in Human Behavior}, 28(3):976 -- 983, 2012.

\bibitem{Quercia@2011}
Daniele Quercia, Michal Kosinski, David Stillwell, and Jon Crowcroft.
\newblock Our twitter profiles, our selves: Predicting personality with
  twitter.
\newblock pages 180--185, 10 2011.

\bibitem{Farnadi@2016}
Golnoosh Farnadi, Geetha Sitaraman, Shanu Sushmita, Fabio Celli, Michal
  Kosinski, David Stillwell, Sergio Davalos, Marie-Francine Moens, and Martine
  De~Cock.
\newblock Computational personality recognition in social media.
\newblock {\em User Modeling and User-Adapted Interaction}, 26, 02 2016.

\bibitem{LolArt}
League of~Legends.
\newblock The art of league of legends.
\newblock \url{https://universe.leagueoflegends.com/en_US/champions/}.

\bibitem{Lankveld@6032007}
G.~{van Lankveld}, P.~{Spronck}, J.~{van den Herik}, and A.~{Arntz}.
\newblock Games as personality profiling tools.
\newblock In {\em 2011 IEEE Conference on Computational Intelligence and Games
  (CIG'11)}, pages 197--202, Aug 2011.

\bibitem{Yee:2006:DMD:1159982.1159988}
Nick Yee.
\newblock The demographics, motivations, and derived experiences of users of
  massively multi-user online graphical environments.
\newblock {\em Presence: Teleoper. Virtual Environ.}, 15(3):309--329, June
  2006.

\bibitem{GOLDBERG1998393}
Lewis~R. Goldberg, Dennis Sweeney, Peter~F. Merenda, and John~Edward Hughes.
\newblock Demographic variables and personality: the effects of gender, age,
  education, and ethnic/racial status on self-descriptions of personality
  attributes.
\newblock {\em Personality and Individual Differences}, 24(3):393 -- 403, 1998.

\bibitem{Griffiths2004}
Mark Griffiths, Mark Davies, and Darren Chappell.
\newblock Demographic factors and playing variables in online computer gaming.
\newblock {\em Cyberpsychology \& behavior : the impact of the Internet,
  multimedia and virtual reality on behavior and society}, 7:479--87, 09 2004.

\bibitem{Suler2004}
John Suler.
\newblock The online disinhibition effect.
\newblock {\em CyberPsychology \& Behavior}, 7(3):321--326, 2004.
\newblock PMID: 15257832.

\bibitem{borders2012relationship}
Joseph~B Borders.
\newblock {\em Relationship between personality and video game preferences.}
\newblock PhD thesis, 2012.

\bibitem{Delhove2018}
Martin Delhove and Tobias Greitemeyer.
\newblock The relationship between video game character preferences and
  aggressive and prosocial personality traits.
\newblock {\em Psychology of Popular Media Culture}, 11 2018.

\bibitem{Yee2011IntrovertedE}
Nick Yee, Nicolas Ducheneaut, Lester Nelson, and Peter Likarish.
\newblock Introverted elves \& conscientious gnomes: the expression of
  personality in world of warcraft.
\newblock In {\em CHI}, 2011.

\bibitem{Graham2013}
Lindsay~T. Graham and Samuel~D. Gosling.
\newblock Personality profiles associated with different motivations for
  playing world of warcraft.
\newblock {\em Cyberpsychology, Behavior, and Social Networking},
  16(3):189--193, 2013.
\newblock PMID: 23438267.

\bibitem{Johnson:2010:PMV:1952222.1952281}
Daniel Johnson and John Gardner.
\newblock Personality, motivation and video games.
\newblock In {\em Proceedings of the 22Nd Conference of the Computer-Human
  Interaction Special Interest Group of Australia on Computer-Human
  Interaction}, OZCHI '10, pages 276--279, New York, NY, USA, 2010. ACM.

\bibitem{Lankveld2009}
Giel Lankveld, Sonny Schreurs, and Pieter Spronck.
\newblock Psychologically verified player modelling.
\newblock pages 12--19, 01 2009.

\bibitem{Peng2008}
Wei Peng, Ming Liu, and Yi~Mou.
\newblock Do aggressive people play violent computer games in a more aggressive
  way? individual difference and idiosyncratic game-playing experience.
\newblock {\em CyberPsychology \& Behavior}, 11(2):157--161, 2008.
\newblock PMID: 18422407.

\bibitem{Park2011}
Jowon Park, Yosep Song, and Ching-I Teng.
\newblock Exploring the links between personality traits and motivations to
  play online games.
\newblock {\em Cyberpsychology, Behavior, and Social Networking},
  14(12):747--751, 2011.
\newblock PMID: 21780935.

\bibitem{Anna2017ICDMW}
A.~Sapienza, H.~Peng, and E.~Ferrara.
\newblock Performance dynamics and success in online games.
\newblock In {\em 2017 IEEE International Conference on Data Mining Workshops
  (ICDMW)}, pages 902--909, Nov 2017.

\bibitem{LOLESPORTS2018}
LOLESPORTS STAFF.
\newblock 2018 mid-season.
\newblock
  \url{https://www.lolesports.com/en_US/articles/2018-mid-season-invitational-numbers}.

\bibitem{Sheer2014}
Ian Sheer.
\newblock Player tally for 'league of legends' surges.
\newblock
  \url{https://blogs.wsj.com/digits/2014/01/27/player-tally-for-league-of-legends-surges/}.

\bibitem{LoLguide}
LoL new~player guide.
\newblock New player guide (lol).
\newblock
  \url{https://na.leagueoflegends.com/en/game-info/get-started/new-player-guide/}.

\bibitem{Gosling2003big-five}
{Samuel D} Gosling, {Peter J.} Rentfrow, and {William B} Swann.
\newblock A very brief measure of the big-five personality domains.
\newblock {\em Journal of Research in Personality}, 37(6):504--528, 1 2003.

\bibitem{Digman1990}
J~M Digman.
\newblock Personality structure: Emergence of the five-factor model.
\newblock {\em Annual Review of Psychology}, 41(1):417--440, 1990.

\bibitem{Cattell1996}
Heather E.~P.~Cattell.
\newblock The original big five: A historical perspective.
\newblock {\em European Review of Applied Psychology/Revue Européenne de
  Psychologie Appliquée}, 46:5--14, 01 1996.

\bibitem{Stoet2010}
Gijsbert Stoet.
\newblock Psytoolkit: A software package for programming psychological
  experiments using linux.
\newblock {\em Behavior Research Methods}, 42(4):1096--1104, Nov 2010.

\bibitem{Stoet2017}
Gijsbert Stoet.
\newblock Psytoolkit: A novel web-based method for running online
  questionnaires and reaction-time experiments.
\newblock {\em Teaching of Psychology}, 44(1):24--31, 2017.

\bibitem{Norman@1963}
Warren T.~Norman.
\newblock Toward an adequate taxonomy of personality attributes: Replicated
  factor structure in peer nomination personality ratings.
\newblock {\em Journal of abnormal and social psychology}, 66:574--83, 07 1963.

\bibitem{JSSv067i01}
Douglas Bates, Martin Mächler, Ben Bolker, and Steve Walker.
\newblock Fitting linear mixed-effects models using lme4.
\newblock {\em Journal of Statistical Software, Articles}, 67(1):1--48, 2015.

\bibitem{Bates2011}
Douglas Bates, Martin Mächler, and Bin Dai.
\newblock {\em lme4: Linear Mixed-Effects Models Using S4 Classes}, volume
  0.999375-33.
\newblock 01 2011.

\bibitem{RCoreTeam}
R~Development Core~Team.
\newblock R: A language and environment for statistical computing.
\newblock {\em R. Found. Stat. Comput.}, 1, 01 2011.

\bibitem{gelman2005}
Andrew Gelman.
\newblock Analysis of variance -- why it is more important than ever.
\newblock {\em Ann. Statist.}, 33(1):1--53, 02 2005.

\bibitem{schwarz1978}
Gideon Schwarz.
\newblock Estimating the dimension of a model.
\newblock {\em Ann. Statist.}, 6(2):461--464, 03 1978.

\bibitem{Robert1995BIC}
Robert~E. Kass and Adrian~E. Raftery.
\newblock Bayes factors.
\newblock {\em Journal of the American Statistical Association},
  90(430):773--795, 1995.

\bibitem{Bean2014}
Dr.~Anthony Bean.
\newblock Video gamers and personality: A five-factor model to understand game
  playing style.
\newblock {\em Psychology of Popular Media Culture}, Online First Publication,
  03 2014.

\bibitem{Poropat2009}
Arthur Poropat.
\newblock A meta-analysis of the five-factor model of personality and academic
  performance.
\newblock {\em Psychological bulletin}, 135:322--38, 04 2009.

\end{thebibliography}


\end{document}